\documentclass[preprint,aps,tightenlines,showpacs,superscriptaddress]{revtex4}
\usepackage[dvips]{graphicx}
\usepackage{bm}
\usepackage{dcolumn}
\usepackage{epsfig}

\begin{document}
\title{Dynamical coupled-channels analysis of $^1\text{H}(e,e^\prime \pi)N$ reactions}
\author{B. Juli\'a-D\'{\i}az}
\affiliation{ Excited Baryon Analysis Center (EBAC), Thomas Jefferson National
Accelerator Facility, Newport News, VA 23606, USA}
\affiliation{Department d'Estructura i Constituents de la Mat\`{e}ria
and Institut de Ci\`{e}ncies del Cosmos,
Universitat de Barcelona, E--08028 Barcelona, Spain}
\author{H. Kamano}
\affiliation{ Excited Baryon Analysis Center (EBAC), Thomas Jefferson National
Accelerator Facility, Newport News, VA 23606, USA}
\author{T.-S. H. Lee}
\affiliation{ Excited Baryon Analysis Center (EBAC), Thomas Jefferson National
Accelerator Facility, Newport News, VA 23606, USA}
\affiliation{Physics Division, Argonne National Laboratory,
Argonne, IL 60439, USA}
\author{A. Matsuyama}
\affiliation{ Excited Baryon Analysis Center (EBAC), Thomas Jefferson National
Accelerator Facility, Newport News, VA 23606, USA}
\affiliation{Department of Physics, Shizuoka University, Shizuoka 422-8529, Japan}
\author{T. Sato}
\affiliation{ Excited Baryon Analysis Center (EBAC), Thomas Jefferson National
Accelerator Facility, Newport News, VA 23606, USA}
\affiliation{Department of Physics, Osaka University, Toyonaka,
Osaka 560-0043, Japan}
\author{N. Suzuki}
\affiliation{ Excited Baryon Analysis Center (EBAC), Thomas Jefferson National
Accelerator Facility, Newport News, VA 23606, USA}
\affiliation{Department of Physics, Osaka University, Toyonaka,
Osaka 560-0043, Japan}

\begin{abstract}
We have performed a dynamical coupled-channels analysis of
available $p(e,e' \pi)N$ data in the region of
 $W \leq $ 1.6 GeV and $Q^2 \leq$ 1.45 (GeV/c)$^2$.
The channels included are $\gamma^* N$, $\pi N$, $\eta N$, and
$\pi\pi N$ which has  $\pi\Delta$, $\rho N$, and $\sigma N$
components.
With the hadronic parameters of the model determined in our
 previous investigations of 
$\pi N\rightarrow \pi N, \pi\pi N$ reactions, we have found that
the available data in the considered  $W \leq $ 1.6 GeV region
can be fitted well by only adjusting 
the  bare $\gamma^* N \rightarrow N^*$  helicity amplitudes 
for  the lowest $N^*$ states in  $P_{33}$, $P_{11}$, $S_{11}$ and
$D_{13}$ partial waves. The sensitivity of the resulting parameters
to the amount of data included in the analysis is investigated.
The importance of coupled-channels 
 effect on the $p(e,e'\pi)N$ cross sections is demonstrated.
The meson cloud effect, as required by the unitarity conditions,
on the $\gamma^* N \rightarrow N^*$  form factors are also examined.
Necessary future developments, both experimentally and theoretically,
are discussed.

\end{abstract}

\maketitle

\section{Introduction}

The electromagnetic parameters characterizing  
the excited nucleons ($ N^*)$, in particular the
$\gamma^* N \rightarrow N^*$ form factors, are important information for
understanding the hadron structure within Quantum Chromodynamics (QCD).
With the  efforts in recent years, as reviewed in Ref.~\cite{bl04},
 the world data 
of $\gamma^* N \rightarrow \Delta (1232)$ form factors
are now considered along with
the electromagnetic nucleon form factors as the benchmark data for
developing hadron structure models and testing predictions from
Lattice QCD calculations (LQCD). The main objective of this work is to
explore the extent to which
the available $p(e,e'\pi)N$ data in $W\leq 1.6$~GeV 
can be used to extract the $\gamma^* N \rightarrow N^*$ form factors
for the $N^*$ states up to the so-called
``second'' resonance region.

We employed a dynamical coupled-channels model 
developed in Refs.~\cite{msl07,jlms07,djlss08,jlmss08,kjlms09}. 
This work is an extension of our analysis~\cite{jlmss08} of
pion photoproduction reactions.
We therefore will only recall equations which are relevant to 
the coupled-channels calculations of $p(e,e'\pi)N$ cross sections.
In the helicity-LSJ mixed-representation where the initial
$\gamma N$ state is specified by its helicities $\lambda_\gamma$ and
$\lambda_N$ and the final $MB$ states by the $(LS)J$ angular momentum
variables, the reaction amplitude of 
$\gamma^*(\vec{q}, Q^2) + N (-\vec{q}) \rightarrow
\pi(\vec{k}) + N (-\vec{k})$ at invariant mass $W$ and momentum transfer
$Q^2=-q^\mu q_\mu = \vec{q}^{\,2}-\omega^2 $ can be written within
a Hamiltonian formulation~\cite{msl07}
 as (suppress the isospin quantum numbers)
\begin{eqnarray}
T^{J}_{LS_N\pi N,\lambda_\gamma\lambda_N}(k,q,W,Q^2) =
{t}^{J}_{L S_N \pi N, \lambda_\gamma\lambda_N}(k, q, W,Q^2)
+t^{R,J}_{LS_N\pi N,\lambda_\gamma\lambda_N}(k, q, W,Q^2)\,,
\label{eq:pw-t}
\end{eqnarray}
where $S_N=1/2$ is the nucleon spin, $W=\omega +E_N(q)$ is the invariant mass
of the $\gamma^* N$ system,  and the non-resonant amplitude is
\begin{eqnarray}
& & {t}^{J}_{L S_N \pi N,\lambda_\gamma\lambda_N}(k,q,W,Q^2)\nonumber \\
& &=
{\it v}^{J}_{L S_N \pi N,\lambda_\gamma\lambda_N}(k,q,Q^2)
+\sum_{M'B'}
\sum_{L^{\prime}S^{\prime}}
\int k^{\prime 2}dk^{\prime }
 {t}^{J}_{L S_N \pi N, L' S'M'B'}(k,k^{\prime},W) \nonumber \\
& & \times G_{M'B'}(k',W)
{\it v}^{J}_{L' S' M'B' , \lambda_\gamma\lambda_N}(k',q,Q^2)\,.
\label{eq:pw-nonr}
\end{eqnarray}
In the above equation, $G_{M'B'}(k',W)$ are  the meson-baryon propagators  for
the channels
 $M'B'= \pi N, \eta N, \pi\Delta, \rho N, \sigma N$. The matrix
elements ${\it v}^{J}_{L S MB,\lambda_\gamma\lambda_N}(k,q,Q^2)$, 
which
describe the $\gamma N \rightarrow MB$ transitions, are calculated
from tree-diagrams of  a set
of phenomenological Lagrangians describing the interactions between
$\gamma$, $\pi$, $\eta$, $\rho$, $\omega$, $\sigma$, $N$, and
$\Delta$(1232) fields. The details are given explicitly in Appendix F
of Ref.~\cite{msl07}. The hadronic non-resonant amplitudes 
${t}^{J}_{L S_N \pi N, L' S'M'B'}(k,k^{\prime},W)$ are generated from
the model constructed from analyzing the data of
$\pi N \rightarrow \pi N, \pi\pi N$ reactions~\cite{jlms07,kjlms09}.

The resonant amplitude in Eq.~(\ref{eq:pw-t}) is
\begin{eqnarray}
t^{R,J}_{LS_N\pi N,\lambda_\gamma\lambda_N}(k, q, W,Q^2) =
 \sum_{N^*_i, N^*_j}
[\bar{\Gamma}^{J}_{N^*_i,LS_N\pi N}(k,W)]^*
D_{i,j}(W)
\bar{\Gamma}^{J}_{N^*_j,\lambda_\gamma\lambda_N}(q,W,Q^2) \,,
\label{eq:pw-r}
\end{eqnarray}
where the dressed $N^*\rightarrow \pi N$ vertex 
$\bar{\Gamma}^{J}_{N^*_i,LS_N\pi N}(k,W)$ and $N^*$ propagator $D_{i,j}(W)$
have been determined and given explicitly 
in Ref.~\cite{jlmss08}. The quantity relevant to our
later discussions is the dressed
$\gamma ^* N \rightarrow N^*$ vertex function
 defined by
\begin{eqnarray}
\bar{\Gamma}^{J}_{N^*,\lambda_\gamma\lambda_N}(q,W,Q^2)
&=&{\Gamma}^{J}_{N^*,\lambda_\gamma\lambda_N }(q,Q^2) \nonumber \\
&+ &\sum_{M'B'}
\sum_{L^{\prime}S^{\prime}}
\int k^{\prime 2}dk^{\prime }
 \bar{\Gamma}^{J}_{N^*,L'S'M'B'}(k',W) G_{M'B'}(k',W)
{\it v}^{J}_{L' S' M'B' , \lambda_\gamma\lambda_N}(k',q,Q^2)\,. \nonumber \\
& &
\label{eq:pw-v}
\end{eqnarray}

\begin{figure}[t]
\centering
\includegraphics[clip,width=12cm,angle=0]{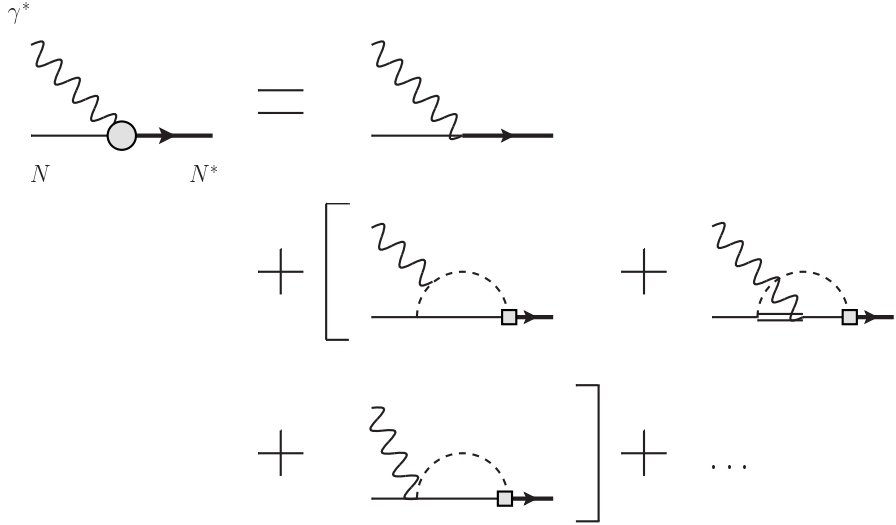}
\caption{Graphical illustration of the contribution to
the $\pi N$ intermediate state to the dressed $\gamma^* N \rightarrow N^*$
vertex defined by Eq.~(\ref{eq:pw-v}).}
\label{fig:d-vertex}
\end{figure}

The second term of Eq.~(\ref{eq:pw-v}) is due to the mechanism 
where
the non-resonant electromagnetic meson production takes place before
the dressed $N^*$ states are formed. This is illustrated in
Fig.~\ref{fig:d-vertex} for the contribution due to
the $M'B' = \pi N$ intermediate state.
Similar to what was defined in Ref.~\cite{sl96,jlss07}, we call this
contribution the {\it meson cloud effect} to define precisely what will
be presented in this paper. We emphasize here that the meson cloud
term in Eq.~(\ref{eq:pw-v}) is the necessary consequence of the unitarity
conditions. How this term and the assumed bare $N^*$ states
are interpreted is obviously model dependent. This issue as well as
the questions concerning the extractions of form factors at
resonance pole positions will be discussed elsewhere, 
and will not be addressed here.

Within the one-photon exchange approximation,
the differential cross sections of pion electroproduction
can be written as
\begin{eqnarray}
\frac{d\sigma^5}{d E_{e'} d\Omega_{e'} d\Omega_\pi^*}
&=&
\Gamma_\gamma
\left[ \sigma_T + \epsilon \sigma_L
 +\sqrt{2\epsilon(1+\epsilon)}\sigma_{LT} \cos \phi_\pi^\ast
\right.
\nonumber \\
& &
\left.
+ \epsilon \sigma_{TT} \cos 2\phi_\pi^\ast
+ h_e\sqrt{2\epsilon(1-\epsilon)}
 \sigma_{LT^\prime}\sin \phi_\pi^\ast
\right].
\label{eq:dcrst-em}
\end{eqnarray}
Here
$\Gamma_\gamma=[\alpha/(2\pi^2 Q^2)]
(E_{e'}/E_e)[|\vec{q}_L|/(1-\epsilon)]$;
$\epsilon$ is defined by the electron scattering angle $\theta_e$
and the photon 3-momentum ${\vec q}_L$ in the laboratory frame
as $\epsilon  =  [1 + 2 (|\vec{q}_L|^2/Q^2)\tan^2 (\theta_{e}/2)]^{-1}$;
$h_e$ is the helicity of the incoming electron;
$\phi_\pi^\ast$ is the angle between
the $\pi$-$N$ plane and the plane of the incoming and outgoing electrons.
The quantities associated with the electrons are defined
in the laboratory frame.
On the other hand, structure functions of $\gamma^* N\to\pi N$ process,
$\sigma_{\alpha}=\sigma_{\alpha}(W,Q^2,\cos\theta_\pi^\ast)$
($\alpha=T,L,LT,TT,LT')$, are defined in the final
$\pi N$ center of mass system.
The formula for calculating $\sigma_{\alpha}$
from the amplitudes defined by Eqs.~(\ref{eq:pw-t})-(\ref{eq:pw-r})
are given in Ref.~\cite{sl09}.

In this first-stage investigation, we only consider the data of
structure functions $\sigma_\alpha$ of $p(e,e'\pi^0)p$~\cite{m98,e1c-pi0ltp}
and $p(e,e'\pi^+)n$~\cite{e1c,e1c-pipltp}
up to $W=1.6$ GeV and $Q^2=1.45$ (GeV/c)$^2$.
The availability of the data in the corresponding $(W,Q^2)$ region
are found in Table~\ref{tab:data-list}. 
The resulting parameters are then confirmed against the original
five-fold differential cross section data~\cite{hallb-site}. 
This procedure could overestimate/underestimate the 
errors of our analysis, but is sufficient for the present 
exploratory investigation.

\begin{table}
\caption{Available structure function data at $Q^2\leq 1.45$ (GeV/c)$^2$.}
\label{tab:data-list}
\begin{ruledtabular}
\begin{tabular}{lll}
$Q^2$ (GeV/c)$^2$ &$\gamma^* p\to\pi^0 p$&$\gamma^* p\to\pi^+ n$\\
\hline
$0.3$  &------&$\sigma_T+\epsilon\sigma_L,~\sigma_{LT},~\sigma_{TT}$~\cite{e1c}\\
$0.4$  &$\sigma_T+\epsilon\sigma_L,~\sigma_{LT},~\sigma_{TT}$~\cite{m98}; $\sigma_{LT'}$~\cite{e1c-pi0ltp} &
        $\sigma_T+\epsilon\sigma_L,~\sigma_{LT},~\sigma_{TT}$~\cite{e1c}; $\sigma_{LT'}$~\cite{e1c-pipltp} \\
$0.5$  &------&$\sigma_T+\epsilon\sigma_L,~\sigma_{LT},~\sigma_{TT}$~\cite{e1c}\footnote{
The data are available up to $W=1.51$~GeV}\\
$0.525$&$\sigma_T+\epsilon\sigma_L,~\sigma_{LT},~\sigma_{TT}$~\cite{m98} &------\\
$0.6$  &------&$\sigma_T+\epsilon\sigma_L,~\sigma_{LT},~\sigma_{TT}$~\cite{e1c}\footnote{
The data are available up to $W=1.41$~GeV}\\
$0.65$ &$\sigma_T+\epsilon\sigma_L,~\sigma_{LT},~\sigma_{TT}$~\cite{m98}; $\sigma_{LT'}$~\cite{e1c-pi0ltp} &$\sigma_{LT'}$~\cite{e1c-pipltp}\\
$0.75$ &$\sigma_T+\epsilon\sigma_L,~\sigma_{LT},~\sigma_{TT}$~\cite{m98}&------\\
$0.9$  &$\sigma_T+\epsilon\sigma_L,~\sigma_{LT},~\sigma_{TT}$~\cite{m98}&------\\
$1.15$ &$\sigma_T+\epsilon\sigma_L,~\sigma_{LT},~\sigma_{TT}$~\cite{m98}&------\\
$1.45$ &$\sigma_T+\epsilon\sigma_L,~\sigma_{LT},~\sigma_{TT}$~\cite{m98}&------
\end{tabular}
\end{ruledtabular}
\end{table}

In section II, we present the results from our analysis.
Discussions on future developments are given in section III.

\section{Analysis and Results}

To proceed, we need to
define the bare $\gamma^\ast N \rightarrow N^*$ vertex
functions $\Gamma^{J}_{N^*,\lambda_\gamma\lambda_N}(q,Q^2)$ 
of Eq.~(\ref{eq:pw-v}).
We parameterize these functions as
\begin{eqnarray}
{\Gamma}^{J}_{N^*,\lambda_\gamma\lambda_N}(q,Q^2)
& =&\frac{1}{(2\pi)^{3/2}}\sqrt{\frac{m_N}{E_N(q)}}\sqrt{\frac{q_R}{|q_0|}}
G_{\lambda}(N^*,Q^2) \delta_{\lambda, (\lambda_\gamma-\lambda_N)}, 
\label{eq:ggn-a} 
\end{eqnarray}
where $q_R$ and $q_0$ are defined by 
$M_{N^*} = q_R+E_N(q_R)$ with $N^*$ mass and
$W = q_0+E_N(q_0)$, respectively, and 
\begin{eqnarray}
G_{\lambda}(N^*,Q^2) &=& A_{\lambda}(N^*,Q^2),~~~\text{for transverse photon}, 
\label{eq:barea}\\
 &=& S_{\lambda}(N^*,Q^2),~~~\,\text{for longitudinal photon}.
\label{eq:bares}
\end{eqnarray}
For later discussions, we
also cast the helicity amplitudes of the
dressed vertex Eq.~(\ref{eq:pw-v}) into the form of
Eq.~(\ref{eq:ggn-a}) with dressed helicity amplitudes
\begin{eqnarray}
\bar{A}_{\lambda}(N^*,Q^2) &=& {A}_{\lambda}(N^*,Q^2) + 
{A}^{\text{m.c.}}_{\lambda}(N^*,Q^2), \label{eq:dressa} \\
\bar{S}_{\lambda}(N^*,Q^2) &=& {S}_{\lambda}(N^*,Q^2) + 
{S}^{\text{m.c.}}_{\lambda}(N^*,Q^2), \label{eq:dresss}
\end{eqnarray}
where ${A}^{\text{m.c.}}_{\lambda}(N^*,Q^2)$ and ${S}^{\text{m.c.}}_{\lambda}(N^*,Q^2)$  are
 due to the meson cloud effect
defined by the second term of Eq.~(\ref{eq:pw-v}).

With the hadronic parameters of
the employed dynamical 
coupled-channels model determined
in analyzing the $\pi N$ reaction data~\cite{jlms07,kjlms09},
the only freedom 
in analyzing the electromagnetic meson production reactions is the
electromagnetic coupling parameters of the model. 
If the parameters
listed in Ref.~\cite{msl07} are used to calculate the non-resonant interaction
${v}^{J}_{L' S' M'B' , \lambda_\gamma\lambda_N}(k',q)$ in Eqs.~(\ref{eq:pw-nonr})
and~(\ref{eq:pw-v}), the only parameters to be determined from the
data of pion electroproduction reactions are
the bare helicity amplitudes
defined by Eq.~(\ref{eq:ggn-a}).
Such a highly constrained
analysis was performed in Ref.~\cite{jlmss08} for pion photoproduction. 
It was found
that the available data of $\gamma p \rightarrow \pi^0 p,~\pi^+ n$
can be fitted reasonably well up to invariant mass $ W\le 1.6$ GeV.
In this work we extend this effort to analyze the pion electroproduction
data in the same $W$ region.

\begin{figure}[t]
\centering
\includegraphics[clip,width=12cm,angle=0]{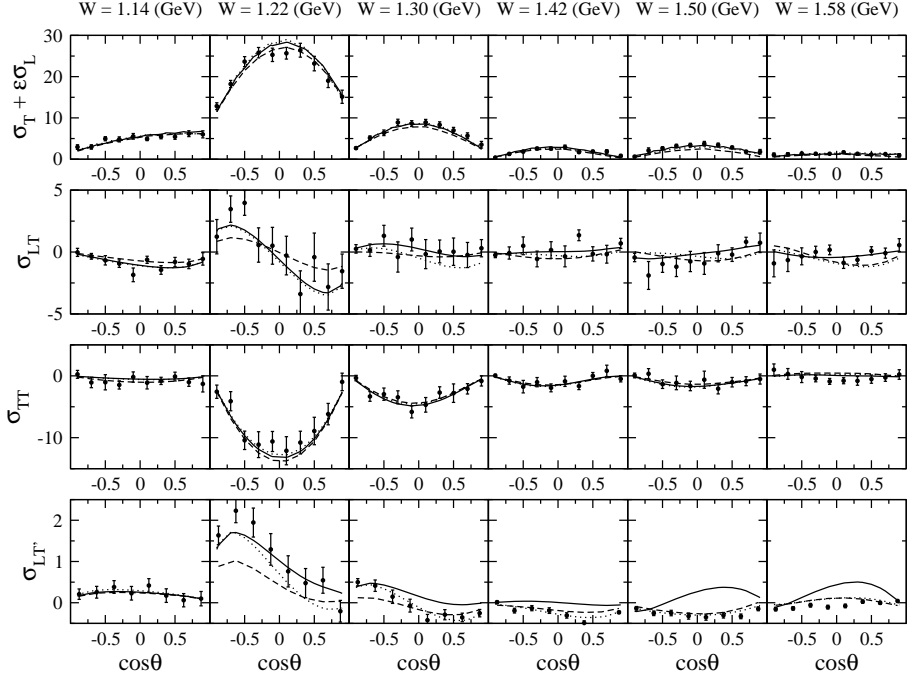}
\caption{Fit to $p(e,e'\pi^0)p$ structure functions at $Q^2=0.4$ (GeV/c)$^2$.
Here $\theta\equiv \theta_\pi^*$.
The solid curves are the results of Fit1,
the dashed curves are of Fit2,
and the dotted curves are of Fit3.
(See text for the description of each fit.)
The data are taken from Refs.~\cite{m98,e1c-pi0ltp}.
}
\label{fig:str-1}
\end{figure}

\begin{figure}[t]
\centering
\includegraphics[clip,width=12cm,angle=0]{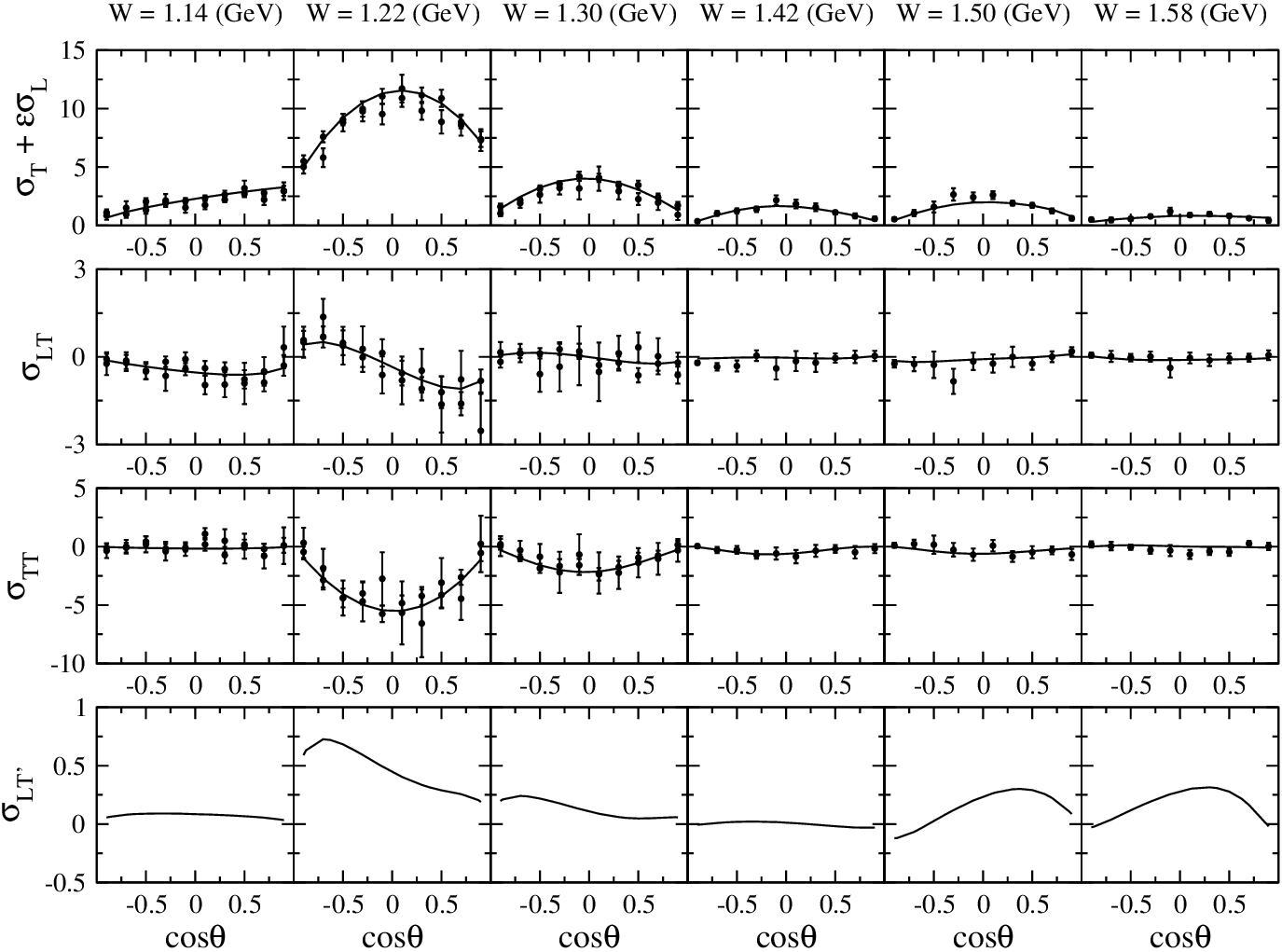}
\caption{Fit to $p(e,e'\pi^0)p$ structure functions at $Q^2=0.9$ (GeV/c)$^2$.
Here $\theta\equiv \theta_\pi^*$.
The data are taken from Ref.~\cite{m98}.
}
\label{fig:str-2}
\end{figure}

\begin{figure}[b]
\centering
\includegraphics[clip,width=12cm,angle=0]{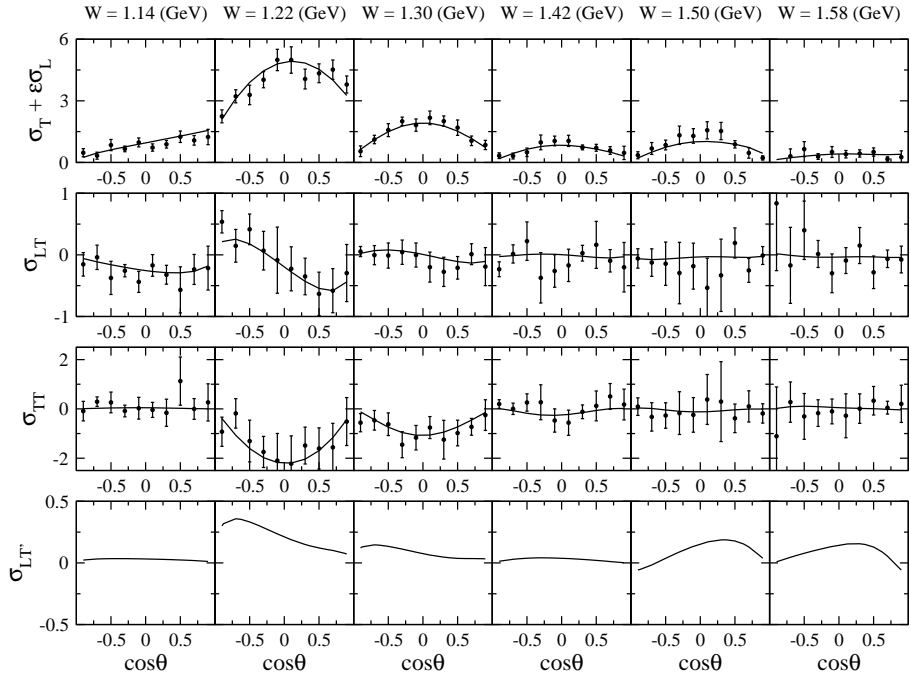}
\caption{Fit to $p(e,e'\pi^0)p$ structure functions at $Q^2=1.45$ (GeV/c)$^2$.
Here $\theta\equiv \theta_\pi^*$.
The data are taken from Ref.~\cite{m98}.
}
\label{fig:str-3}
\end{figure}

\begin{figure}[th]
\centering
\includegraphics[clip,width=8cm,angle=0]{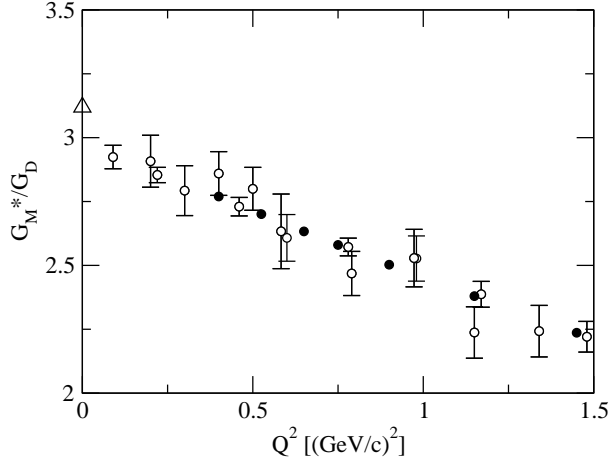}
\caption{$G_M^\ast$ normalized by the dipole factor
$G_D=[1+Q^2/0.71\text{(GeV/c)}^2]^{-2}$.
The solid black circles at $Q^2 > 0$ are our fit to 
the values extracted from previous analyses
(those values are taken from Ref.~\cite{jlmss08}).
The triangle at $Q^2=0$ is from our
photoproduction analysis~\cite{jlmss08}.
}
\label{fig:gm-delta}
\end{figure}

We first try to fix the bare helicity amplitudes by fitting to
the data of $\sigma_T+\epsilon \sigma_L$, $\sigma_{LT}$, and
$\sigma_{TT}$ of $p(e,e'\pi^0)p$
in Ref.~\cite{m98} which covers almost all $(W,Q^2)$ region
we are considering (see Table.~\ref{tab:data-list}).
In a purely phenomenological approach, we first vary 
all of the helicity amplitudes of 16 bare $N^*$ states, considered
in analyzing the $\pi N \rightarrow \pi N,\pi \pi N$ 
data~\cite{jlms07,kjlms09}, in the fits to the data.
It turns out that only the helicity amplitudes of the first $N^*$ states
in $S_{11}$, $P_{11}$, $P_{33}$ and $D_{13}$
are relevant in the considered $W \leq$ 1.6 GeV. 
Thus in this paper only the bare helicity amplitudes associated with those
four bare $N^*$ states (total 10 parameters) are varied in the fit and
other bare helicity amplitudes are set to zero.
The numerical fit is performed at each $Q^2$ independently, using
the MINUIT library.

The results of our fits
are the solid curves in the top three rows of
Figs.~\ref{fig:str-1}-\ref{fig:str-3}. 
Clearly our results from this fit agree with the data well.
We obtain similar quality of fits to the data of Ref.~\cite{m98}
at other $Q^2$ values listed in Table.~\ref{tab:data-list}.
We have also used the magnetic $M1$ form factor
of $\gamma^\ast N\to \Delta(1232)$ 
extracted from previous analyses as data for fitting.
The results are shown in Fig.~\ref{fig:gm-delta}. 
We refer the results of this fit to as ``Fit1''.

\begin{figure}[t]
\centering
\includegraphics[clip,width=12cm,angle=0]{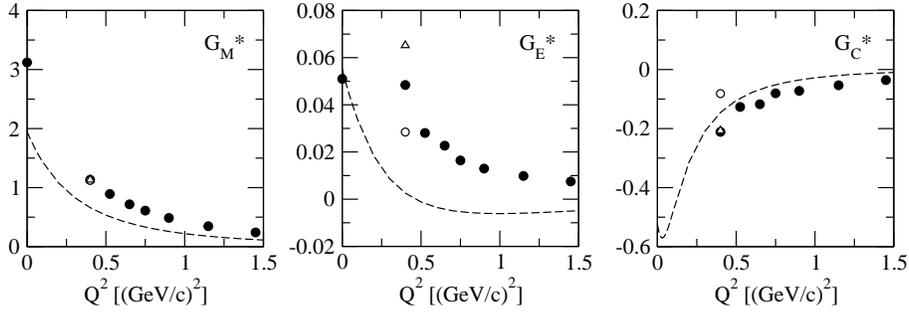}
\caption{The $\gamma^\ast N\to \Delta(1232)$ form factors.
Solid points are from Fit1; dashed curves are the meson cloud contribution.
Open circles and triangles at $Q^2=0.4$ (GeV/c)$^2$
are from Fit2 and Fit3, respectively.
The three points are almost overlapped in $G_M^\ast$.
The solid point at $Q^2=0$ is obtained 
from the photoproduction reaction analysis 
in Ref.~\cite{jlmss08}.
}
\label{fig:gmgegc}
\end{figure}

\begin{figure}[t]
\centering
\includegraphics[clip,width=12cm,angle=0]{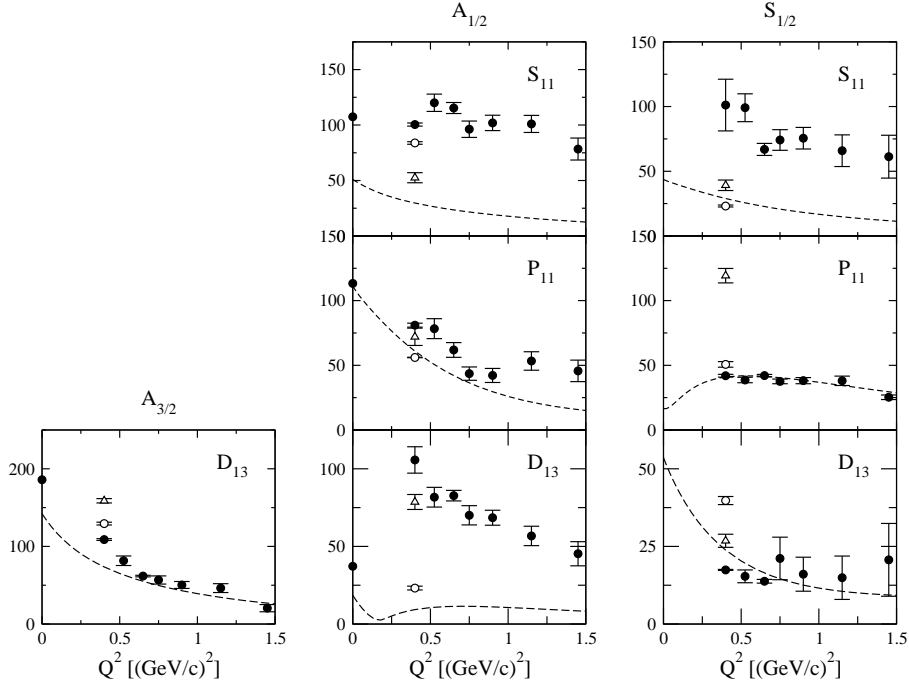}
\caption{
Extracted helicity amplitudes for
$S_{11}$  at $W=1535$ MeV (upper panels),
$P_{11}$  at $W=1440$ MeV (middle panels),
and $D_{13}$  at $W=1520$ MeV (lower panels).
The meaning of each point and curve is same as in Fig.~\ref{fig:gmgegc}.
}
\label{fig:hel-s11p11d13}
\end{figure}

In Fig.~\ref{fig:gmgegc}, we present
the $G_M^\ast$, $G_E^\ast$, and $G_C^\ast$ form factors of 
$\gamma^\ast N\to \Delta(1232)$ transition obtained from 
Fit1 (solid points).
In the same figure, we also show the meson cloud effect
in the form factors. 
Within our model, it has a significant contribution at low $Q^2$,
but rapidly decreases as $Q^2$ increases, particularly for
$G_E^\ast$ and $G_C^\ast$.
These results are similar to the previous findings~\cite{sl01,jlss07}.

The helicity amplitudes of
$S_{11}$, $P_{11}$, and $D_{13}$ resulting from Fit1 are shown in 
Fig.~\ref{fig:hel-s11p11d13}.
The solid circles are the absolute magnitude of the dressed helicity amplitudes
(\ref{eq:dressa}) and~(\ref{eq:dresss}).
The errors there are assigned by MIGRAD in the MINUIT library.
More detailed analysis of the errors is perhaps needed, but will not
be addressed here.
The meson cloud effect (dashed curves), as
defined by $A^{\text{m.c.}}_\lambda$ and $S^{\text{m.c.}}_\lambda$
of Eqs.~(\ref{eq:dressa}) and~(\ref{eq:dresss}) and
calculated from the second term of Eq.~(\ref{eq:pw-v}),
are the necessary consequence of 
the unitarity conditions. 
They do not include the bare helicity term determined here 
and are already fixed in the photoproduction analysis~\cite{jlmss08}.
Within our model (and within Fit1), 
the meson cloud contribution is relatively small
in $S_{11}$ and $A_{1/2}$ of $D_{13}$ even in the low $Q^2$ region.

Here we note that our helicity amplitudes defined in 
Eqs.~(\ref{eq:dressa}) and~(\ref{eq:dresss})
are different from the commonly used convention, say 
$A^{\text{cnv}}_\lambda$ and $S^{\text{cnv}}_\lambda$,
which are obtained from the imaginary part of the $\gamma^\ast N\to \pi N$
multipole amplitudes~\cite{abl08}.
This definition leads to helicity amplitudes which are real,
while our dressed amplitudes are complex.
It was shown in Ref.~\cite{sl01} that
for the $\Delta(1232)$ resonance our dressed helicity amplitudes 
(\ref{eq:dressa}) and (\ref{eq:dresss})
can be reduced to $A^{\text{cnv}}_\lambda $ and $S^{\text{cnv}}_\lambda $,
if we replace the Green function $G_{\pi N}$ with its principal value
in all loop integrals appearing in the calculation.
However, such reduction is not so trivial for higher resonance states
because the unstable $\pi\Delta,\rho N,\sigma N$ channels open,
and thus the direct comparison of the helicity amplitudes
from other analyses becomes unclear.

\begin{figure}[h]
\centering
\includegraphics[clip,width=12cm,angle=0]{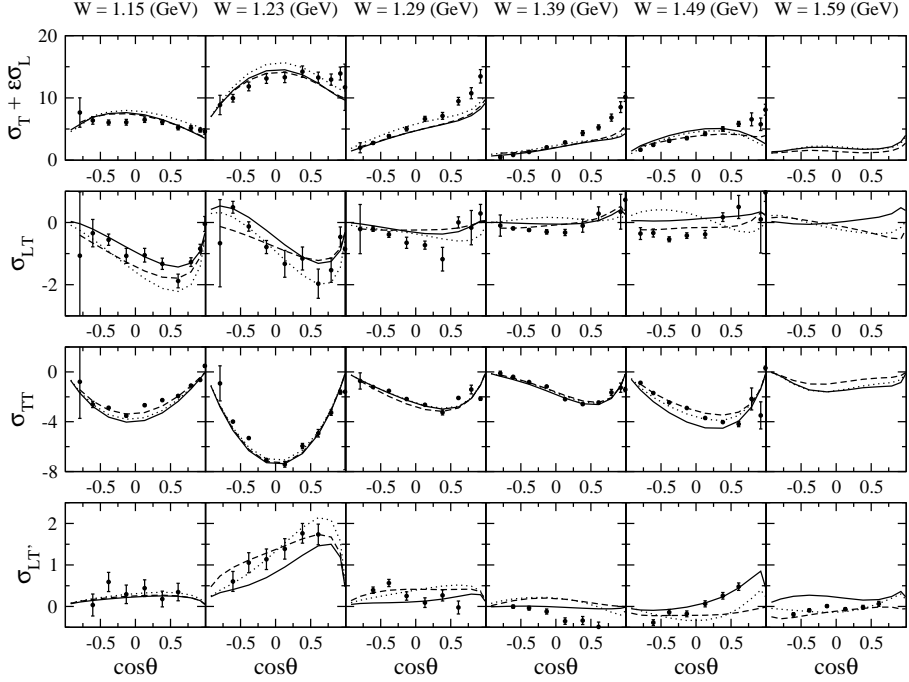}
\caption{
Structure functions of $p(e,e'\pi^+)n$ at $Q^2=0.4$ (GeV/c)$^2$.
Here $\theta\equiv\theta_\pi^*$.
The solid curves are the results of Fit1,
the dashed curves are of Fit2,
and the dotted curves are of Fit3.
(See text for the description of each fit.)
As for the $\sigma_{LT'}$, results at $W=1.14,1.22,1.3,1.38,1.5,1.58$ GeV
(from left to right of the bottom row) are shown, in which the data
are available.
The data in the figure are taken from Ref.~\cite{e1c,e1c-pipltp}.
}
\label{fig:str-4}
\end{figure}

At $Q^2=0.4$ (GeV/c)$^2$, the data of all structure functions
both for $p(e,e'\pi^0)p$ and $p(e,e'\pi^+)n$ 
are available as seen in Table.~\ref{tab:data-list}. 
To see the sensitivity of the resulting helicity amplitudes to the
amount of the data included in the fits,
we further carry out two fits at this $Q^2$, 
referred to as Fit2 and Fit3, respectively.
Fit2 (Fit3) further includes the data of 
Refs.~\cite{e1c-pi0ltp,e1c,e1c-pipltp} (Ref.~\cite{e1c-pi0ltp})
in the fit in addition to those of Ref.~\cite{m98} which are used in Fit1.
This means that Fit2 includes all available data
both from $p(e,e'\pi^0)p$ and $p(e,e'\pi^+)n$, whereas
Fit3 includes the same data but from $p(e,e'\pi^0)p$ only.
The results of each fit are the dashed and dotted curves in 
Fig.~\ref{fig:str-1} for
$p(e,e'\pi^0)p$  and Fig.~\ref{fig:str-4} for $p(e,e'\pi^+)n$,
respectively.

The resulting bare helicity amplitudes are listed in the third (Fit2)
and fourth (Fit3) columns of Table II and compare with that from
Fit1. 
The corresponding change in the $\gamma N\to \Delta(1232)$ form
factors and the dressed helicity amplitudes are also shown
as open circles and triangles
in Figs.~\ref{fig:gmgegc} and~\ref{fig:hel-s11p11d13}.
A significant change among the three different fits
is observed in most of the results except $G^\ast_{M}$ in $P_{33}$.
This indicates that fitting the data listed in Table~\ref{tab:data-list}
 are far from
sufficient to pin down the $\gamma^* N \rightarrow N^*$ 
transition form factors up to $Q^2=1.45$ (GeV/c)$^2$. 
It clearly indicates the importance of
obtaining data from complete or over-complete measurements of most, if 
not all, of the independent $p(e,e'\pi)N$ 
polarization observables. 
Such measurements were made by Kelly
et al.~\cite{kelly} in the $\Delta$ (1232) region and
will be  performed at JLab for wide ranges of W and $Q^2$
in the next few years~\cite{bl04}.

\begin{table}
\label{tab:bare-hel}

\caption{Ambiguity of resulting bare helicity amplitudes 
[the results are at $Q^2=0.4$ (GeV/c)$^2$].
The errors are assigned by MIGRAD in the MINUIT library.}
\begin{ruledtabular}
\begin{tabular}{crrrr}
&Fit1&Fit2 & Fit3 \\
&(Ref.~\cite{m98} data)&(Refs.~\cite{m98,e1c-pi0ltp,e1c,e1c-pipltp} data)&
(Refs.~\cite{m98,e1c-pi0ltp} data)\\
\hline
$S_{11}~A_{1/2}$ &$100.80 \pm 1.46 $ &$83.25  \pm 1.21$ &$48.29\pm 5.46$ \\
$S_{11}~S_{1/2}$ &$-119.30\pm 20.41$ &$-9.85  \pm 1.69$ &$-53.53\pm 4.75$ \\
$P_{11}~A_{1/2}$ &$33.18  \pm 2.11$ &$-15.68 \pm 1.00$ &$20.17\pm 10.37$\\   
$P_{11}~S_{1/2}$ &$37.29  \pm 2.26$ &$52.23  \pm 3.16$ &$131.00\pm 5.87$\\
$P_{33}~A_{3/2}$ &$-146.00\pm 0.60$ &$-137.50\pm 0.56$ &$-150.80\pm 1.03$\\
$P_{33}~A_{1/2}$ &$-54.47 \pm 0.61$ &$-62.57 \pm 0.69$ &$-46.29 \pm 1.73$\\ 
$P_{33}~S_{1/2}$ &$7.85   \pm 1.25$ &$-7.66  \pm 1.22$ &$7.34   \pm 1.69$\\ 
$D_{13}~A_{3/2}$ &$-44.01 \pm 1.31$ &$-67.01 \pm 1.99$ &$-98.63\pm 2.92$\\  
$D_{13}~A_{1/2}$ &$97.11  \pm 8.51$ &$14.34  \pm 1.26$ &$70.02\pm 4.83$\\ 
$D_{13}~S_{1/2}$ &$-18.35 \pm 1.37$ &$19.43  \pm 1.45$ &$4.11\pm 2.76$\\ 
\end{tabular}
\end{ruledtabular}
\end{table}

\begin{figure}[ht]
\centering
\includegraphics[clip,height=12cm,angle=0]{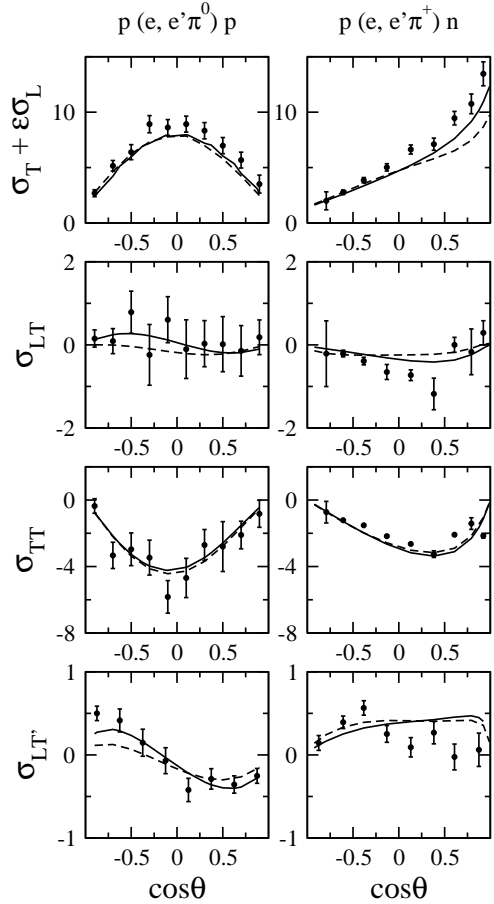}
\caption{
Contribution of the $S_{31}$ and $P_{13}$ helicity amplitudes
at $Q^2=0.4$ (GeV/c)$^2$.
The left (right) panels are the structure functions 
of $p(e,e'\pi^0)p$ [$p(e,e'\pi^+)n$] reaction at
$W=1.3$ GeV ($W=1.29$ GeV).
Solid (dashed) curves are the results with (without) 
nonzero $S_{31}$ and $P_{13}$ bare helicity amplitudes. 
The parameters of Fit2 are used for the 
$S_{11}$, $P_{11}$, $P_{33}$, and $D_{13}$ helicity amplitudes
in both curves.
The data are from Refs.~\cite{m98,e1c-pi0ltp,e1c,e1c-pipltp}.
}
\label{fig:s31p13}
\end{figure}

It has been seen in Fig.~\ref{fig:str-4} that all of our current fits
underestimate $\sigma_T$ of $p(e,e'\pi^+)n$ at forward angles.
We find that this can be improved by further varying 
the $S_{31}$ and $P_{13}$ bare helicity amplitudes 
within their reasonable range.
In Fig.~\ref{fig:s31p13}, the results with 
the nonzero $S_{31}$ and $P_{13}$ bare helicity amplitudes (solid curves)
are compared with the results without varying those amplitudes (dashed curves).
The resulting values of the bare helicity amplitudes are 
$(A_{1/2}^{S_{31}},S_{1/2}^{S_{31}})=(121.6,59.6)$ and
$(A_{3/2}^{P_{13}},A_{1/2}^{P_{13}},S_{1/2}^{P_{13}})=(-73.2,-42.9,41.5)$.
The parameters of Fit2 are used for 
$S_{11}$, $P_{11}$, $P_{33}$, and $D_{13}$ in both curves.
In the figure we have just shown the results at $W\sim 1.3$ GeV.
We confirm that the same consequence is obtained also at other $W$,
and find that the $P_{13}$ ($S_{31}$) 
has contributions mainly at low (high) $W$.
We also find that the inclusion of the bare $S_{31}$ and $P_{13}$
helicity amplitudes does not change other structure functions than 
$\sigma_T$ of $p(e,e'\pi^+)n$ 
(at most, most of the change is within the error).
This indicates that those two helicity amplitudes are rather relevant
to $p(e,e'\pi^+)n$, but not to $p(e,e'\pi^0)p$.
As shown in Table~\ref{tab:data-list}, however,
no enough data is currently available 
for $p(e,e'\pi^+)n$ above $Q^2=0.4$ (GeV/c)$^2$.
The data both of the 
$p(e,e'\pi^0)p$ and $p(e,e'\pi^+)n$ 
at same $Q^2$ values are desirable
to pin down the $Q^2$ dependence of the
$S_{31}$ and $P_{13}$ helicity amplitudes.

\begin{figure}[h]
\centering
\includegraphics[clip,width=10cm,angle=0]{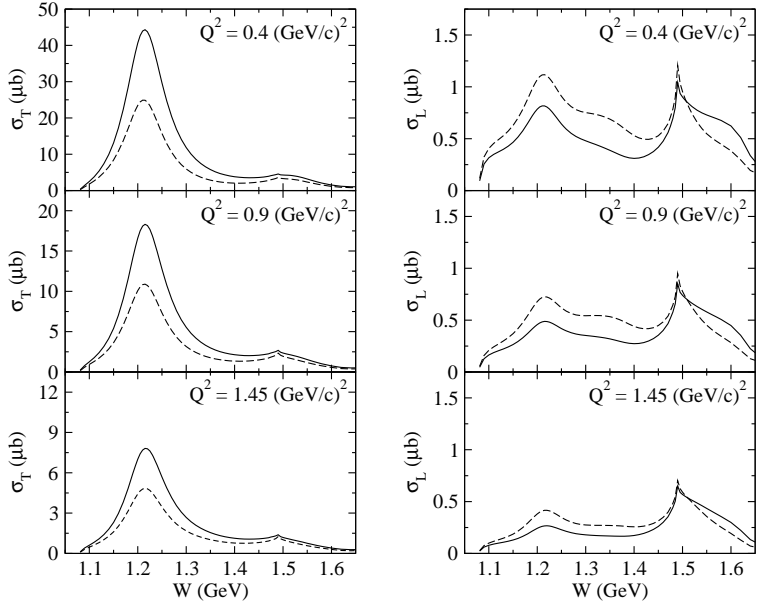}
\caption{
Coupled-channels effect on the integrated structure functions $\sigma_T(W)$
and $\sigma_L(W)$ for $Q^2=0.4,~0.9,~1.45 $ (GeV/c)$^2$ for
$p(e,e'\pi^0)p$ reactions. 
The solid curves are the full results calculated with 
the bare helicity amplitudes of Fit1.
The dashed curves are the same as solid curves but only the $\pi N$ loop is
taken in the $M'B'$ summation in Eqs.~(\ref{eq:pw-nonr}) 
and~(\ref{eq:pw-v}).
 }
\label{fig:cc-pi0p}
\end{figure}

We now turn to show the coupled-channels effect. 
In Fig.~\ref{fig:cc-pi0p}, we see that when only the $\pi N$
intermediate state is kept in the $M'B'$ summation of
the non-resonant amplitude [Eq.~(\ref{eq:pw-nonr})] and
the dressed $\gamma^* N \to N^*$ vertices [Eq.~(\ref{eq:pw-v})], 
the predicted total transverse and longitudinal cross sections
$\sigma_T$ and $\sigma_L$ of $p(e,e'\pi^0)p$ are 
changed from the solid to dashed curves.
This corresponds to only examining
the coupled-channels effect on the electromagnetic ($Q^2$-dependent) part
in the $\gamma^\ast N\to \pi N$ amplitude.
All coupled-channels effects on the non-electromagnetic interactions
are kept in the calculations.
We find that the coupled-channels effect 
tends to decrease when $Q^2$ increases.
This is rather clearly seen in $\sigma_T$.
In particular, the coupled-channels effect on $\sigma_T$ at high 
$W\sim 1.5$ GeV is small ($10$-$20$\%) already at $Q^2 = 0.4 $ (GeV/c)$^2$.
(The effect is about $30$-$40$\% at $Q^2=0$~\cite{jlmss08}.)
This is understood as follows.
In Eq.~(\ref{eq:pw-r}) we can further split the resonant amplitude $t^R$ 
as $t^R = t^R_{\text{bare}} + t^R_{\text{m.c.}}$,
where $t^R_{\text{bare}}$ and $ t^R_{\text{m.c.}}$
are the same as $t^R$ but replacing 
$\bar\Gamma^J_{N^\ast,\lambda_\gamma\lambda_N}$
with its bare part $\Gamma^J_{N^\ast,\lambda_\gamma\lambda_N}$
and meson cloud part [the second term of Eq.~(\ref{eq:pw-v})], respectively.
The coupled-channels effect shown in Fig.~\ref{fig:cc-pi0p}
comes from $t^{J}_{LS_{N}\pi N,\lambda_\gamma\lambda_N}$ and $t^R_{\text{m.c.}}$.
We have found that the relative importance of 
the coupled-channels effect
in each part remains the same for increasing $Q^2$.
However, the contribution of non-resonant mechanisms both on
$t^{J}_{LS_{N}\pi N,\lambda_\gamma\lambda_N}$ and $t^R_{\text{m.c.}}$
to the structure functions decreases for higher $Q^2$
compared with $t^R_{\text{bare}}$.
This explains the smaller coupled-channels effect
compared with the photoproduction reactions~\cite{jlmss08}.
The decreasing non-resonant interaction at higher $Q^2$ is due to 
its long range nature, thus indicating that
higher $Q^2$ reactions provide a clearer probe of $N^\ast$.
We obtain similar results also for $p(e,e'\pi^+)n$.

It is noted, however, that the above argument does not mean 
coupled-channels effect is negligible in the full $\gamma^\ast N\to \pi N$
reaction process.
In the above analysis we kept the coupled-channels effect 
on the hadronic non-resonant amplitudes, the strong $N^\ast$ vertices, 
and the $N^\ast$ self-energy, which are $Q^2$-independent
and remain important irrespective of $Q^2$.
We have found in the previous analyses~\cite{jlms07,kjlms09} 
that the coupled-channels effect on them is significant in all energy region
up to $W= 2$ GeV.

\begin{figure}[ht]
\centering
\includegraphics[clip,width=10cm,angle=0]{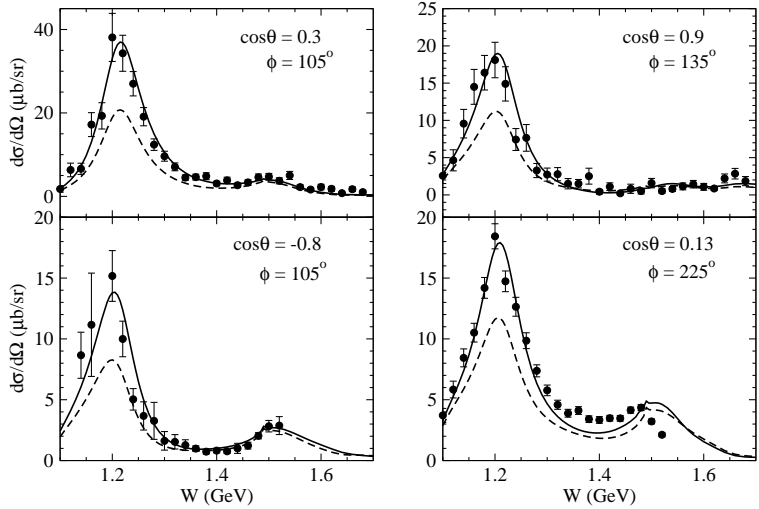}
\caption{
Coupled-channels effect on the five-fold differential cross sections 
$\Gamma_\gamma^{-1}[d\sigma^5/(d E_{e'}d\Omega_{e'}d\Omega_\pi^*)]$ 
of $p(e,e'\pi^0)p$ (upper panels) and $p(e,e'\pi^+)n$ (lower panels)
at $Q^2=0.4$ (GeV/c)$^2$.
Here $\theta\equiv\theta_\pi^*$ and $\phi\equiv\phi_\pi^*$.
The solid curves are the full results calculated with 
the bare helicity amplitudes of Fit1.
The dashed curves are the same as the solid curves but only 
the $\pi N$ loop is taken in 
the $M'B'$ summation in Eqs.~(\ref{eq:pw-nonr}) and~(\ref{eq:pw-v}).
The data are taken from Ref.~\cite{hallb-site}}.
\label{fig:5dim}
\end{figure}

In Fig.~\ref{fig:5dim}, we show the coupled-channels effect on the five-fold 
differential cross section defined by Eq.~(\ref{eq:dcrst-em}).
The coupled-channels effects are significant at low $W$,
whereas they are small at high $W$.
This is consistent with the above discussions because
the five-fold differential cross sections are dominated by $\sigma_T$.
Here we also see that our full results (solid curves) are 
in good agreement with the original data,
although we performed the fits by using the structure function data listed in
Table~\ref{tab:data-list}.

\section{Summary and outlook}

In this work we have explored how the available $p(e,e^\prime \pi)N$ data
can be used to determine the $\gamma^* N \rightarrow N^*$ transition
form factors within a dynamical 
coupled-channels models~\cite{msl07,jlms07,jlmss08,djlss08,kjlms09}.
Within the available data,
the $\gamma^* N \rightarrow N^*$ bare helicity amplitudes of 
the first $N^*$ states in  $S_{11}$, $P_{11}$, $P_{33}$ and $D_{13}$
can be determined in the considered energy region $W \leq$ 1.6 GeV.
We further observe that some of these parameters can not
be determined  well. The uncertainties could be due to
the limitation that only data of
4 out of 11
independent $p(e,e'\pi)N$ observables are available for our
analysis. Clearly, 
the data from the forthcoming measurements of double and triple polarization
observables at JLab
will be highly desirable to make progress.

Also, it was found that the underestimation of 
the $\sigma_T$ of $p(e,e'\pi^+)n$ at forward angles
can be improved by further considering
the $S_{31}$ and $P_{13}$ bare helicity amplitudes.
Furthermore, these amplitudes can have relevant contribution
to $p(e,e'\pi^+)n$, but not to $p(e,e'\pi^0)p$.
The $p(e,e'\pi^+)n$ data of wide $Q^2$ region
as well as $p(e,e'\pi^0)p$ seem necessary for
determining the $Q^2$ dependence of the
$S_{31}$ and $P_{13}$ helicity amplitudes.

For testing theoretical predictions from hadron structure calculations
such as LQCD, the quantities of interest are the residues of the
$\gamma^* N \rightarrow \pi N$ amplitudes, defined by 
Eqs.~(\ref{eq:pw-t})-(\ref{eq:pw-v}), at the 
corresponding resonance poles.
If the resonance poles are associated with the amplitude
$t^{R,J}_{LS_N\pi N,\lambda_\gamma\lambda_N}(k, q, W,Q^2)$ of
Eq.~(\ref{eq:pw-r}), the extracted residues are directly related to
the dressed form factors $\bar{\Gamma}^{J}_{N^*,L'S'M'B'}(k',W)$.
An analytic continuation method for extracting these information
has been developed~\cite{ssl09}, and our results along with
other hadronic properties associated nucleon resonances
will be published elsewhere. Here we only mention that
the extracted form factors are complex and
some investigations are needed to see how
they can be compared with the helicity amplitudes, which are real numbers,
listed by PDG~\cite{pdg}.
In a Hamiltonian formulation as taken in our dynamical approach, the
physical meanings of poles and residues  are well defined in
textbooks~\cite{goldberger,feshbach}.

\begin{acknowledgments}
We would like to thank Dr. K. Park for sending the structure function data
from CLAS.
This work is supported by 
the U.S. Department of Energy, Office of Nuclear Physics Division, under 
contract No. DE-AC02-06CH11357, and Contract No. DE-AC05-06OR23177 
under which Jefferson Science Associates operates Jefferson Lab,
by the Japan Society for the Promotion of Science,
Grant-in-Aid for Scientific Research(C) 20540270, 
and by a CPAN Consolider INGENIO CSD 2007-0042 contract 
and Grants No. FIS2008-1661 (Spain).
This work used resources of the National Energy Research Scientific
Computing Center which is supported by the Office of Science of the 
U.S. Department of Energy under Contract No. DE-AC02-05CH11231.
\end{acknowledgments}

\end{document}